\documentstyle[times,psfig,mathrsfs]{mn}

\newif\ifAMStwofonts
\AMStwofontstrue

\ifoldfss
  \ifCUPmtlplainloaded \else
    \NewTextAlphabet{textbfit} {cmbxti10} {}
    \NewTextAlphabet{textbfss} {cmssbx10} {}
    \NewMathAlphabet{mathbfit} {cmbxti10} {} 
    \NewMathAlphabet{mathbfss} {cmssbx10} {} 
  \fi
  \ifAMStwofonts
    \ifCUPmtlplainloaded \else
      \NewSymbolFont{upmath} {eurm10}
      \NewSymbolFont{AMSa} {msam10}
      \NewMathSymbol{\upi}     {0}{upmath}{19}
      \NewMathSymbol{\umu}     {0}{upmath}{16}
      \NewMathSymbol{\upartial}{0}{upmath}{40}
      \NewMathSymbol{\leqslant}{3}{AMSa}{36}
      \NewMathSymbol{\geqslant}{3}{AMSa}{3E}

    \fi
  \fi
\fi 

\ifnfssone
  \newmathalphabet{\mathit}
  \addtoversion{normal}{\mathit}{cmr}{m}{it}
  \addtoversion{bold}{\mathit}{cmr}{bx}{it}
  \newmathalphabet{\mathbfit} 
  \addtoversion{normal}{\mathbfit}{cmr}{bx}{it}
  \addtoversion{bold}{\mathbfit}{cmr}{bx}{it}
  \newmathalphabet{\mathbfss} 
  \addtoversion{normal}{\mathbfss}{cmss}{bx}{n}
  \addtoversion{bold}{\mathbfss}{cmss}{bx}{n}
  \ifAMStwofonts
    \ifCUPmtlplainloaded \else
      \UseAMStwoboldmath
      \makeatletter
      \new@mathgroup\upmath@group
      \define@mathgroup\mv@normal\upmath@group{eur}{m}{n}
      \define@mathgroup\mv@bold\upmath@group{eur}{b}{n}
      \edef\UPM{\hexnumber\upmath@group}
      \new@mathgroup\amsa@group
      \define@mathgroup\mv@normal\amsa@group{msa}{m}{n}
      \define@mathgroup\mv@bold\amsa@group{msa}{m}{n}
      \edef\AMSa{\hexnumber\amsa@group}
      \makeatother
      \mathchardef\upi="0\UPM19
      \mathchardef\umu="0\UPM16
      \mathchardef\upartial="0\UPM40
      \mathchardef\leqslant="3\AMSa36
      \mathchardef\geqslant="3\AMSa3E
    \fi
  \fi
\fi 

\ifnfsstwo
  \DeclareMathAlphabet{\mathbfit}{OT1}{cmr}{bx}{it}
  \SetMathAlphabet\mathbfit{bold}{OT1}{cmr}{bx}{it}
  \DeclareMathAlphabet{\mathbfss}{OT1}{cmss}{bx}{n}
  \SetMathAlphabet\mathbfss{bold}{OT1}{cmss}{bx}{n}
  \ifAMStwofonts
    \ifCUPmtlplainloaded \else
      \DeclareSymbolFont{UPM}{U}{eur}{m}{n}
      \SetSymbolFont{UPM}{bold}{U}{eur}{b}{n}
      \DeclareSymbolFont{AMSa}{U}{msa}{m}{n}
      \DeclareMathSymbol{\upi}{0}{UPM}{"19}
      \DeclareMathSymbol{\umu}{0}{UPM}{"16}
      \DeclareMathSymbol{\upartial}{0}{UPM}{"40}
      \DeclareMathSymbol{\leqslant}{3}{AMSa}{"36}
      \DeclareMathSymbol{\geqslant}{3}{AMSa}{"3E}
    \fi
  \fi
\fi 

\ifCUPmtlplainloaded \else
  \ifAMStwofonts \else 
    \def\upi{\pi}
    \def\umu{\mu}
    \def\upartial{\partial}
  \fi
\fi

   \title[Copper evolution]{Contrasting copper evolution in 
          ${\bmath{\omega}}$\,Centauri and the Milky Way}
   \author[D. Romano \& F. Matteucci]{Donatella Romano$^{1}$\thanks{E-mail: 
           donatella.romano@oabo.inaf.it} and Francesca Matteucci$^{2}$\\
           $^{1}$INAF\,--\,Osservatorio Astronomico di Bologna,
                 Via Ranzani 1, I-40127 Bologna, Italy\\
           $^{2}$Dipartimento di Astronomia, Universit\`a di Trieste,
                 Via Tiepolo 11, I-34131 Trieste, Italy}
   
\begin{document}

     \date{Accepted 2007 March 29. Received 2007 March 27; in original form 
           2007 March 5}

     \pagerange{\pageref{firstpage}--\pageref{lastpage}} \pubyear{2007}

     \maketitle

     \label{firstpage}


   \begin{abstract}
     Despite the many studies on stellar nucleosynthesis published so far, the 
     scenario for the production of Cu in stars remains elusive. In 
     particular, it is still debated whether copper originates mostly in 
     massive stars or type Ia supernovae. To answer this question, we compute 
     self-consistent chemical evolution models taking into account the results 
     of updated stellar nucleosynthesis. By contrasting copper evolution in 
     $\omega$\,Cen and the Milky Way, we end up with a picture where massive 
     stars are the major responsible for the production of Cu in $\omega$\,Cen 
     as well as the Galactic disc.
   \end{abstract}

   \begin{keywords}
     Galaxy: evolution -- globular clusters: individual: $\omega$ Centauri -- 
     nuclear reactions, nucleosynthesis, abundances -- stars: abundances.
   \end{keywords}


   \section{Introduction}
   \label{sec:int}

   Almost two decades ago, Sneden \& Crocker (1988) and Sneden, Gratton \& 
   Crocker (1991) discussed the possible mechanisms for the synthesis of Cu in 
   stars in the light of their high-resolution observations (showing a decline 
   in [Cu/Fe] with decreasing [Fe/H] in Galactic stars) and favoured a 
   secondary origin for Cu through the weak component of the \emph{s-}process 
   in massive stars.

   The \emph{s-}process has long been thought to contribute to the production 
   of Cu in stars (see, e.g., Burbidge et al. 1957). The main 
   \emph{s-}process, operating in low-mass asymptotic giant branch (AGB) 
   stars, does not contribute more than 5 per cent to the synthesis of Cu 
   isotopes (Raiteri et al. 1993; Travaglio et al. 2004) and will be neglected 
   hereinafter. The weak \emph{s-}process in massive stars depends on 
   metallicity. This mechanism, in fact, requires neutron fluxes originating 
   mainly from the reaction $^{22}$Ne\,($\alpha$, n)\,$^{25}$Mg, where the 
   large abundance of $^{22}$Ne during core He burning derives from the 
   original CNO nuclei trasmuted into $^{14}$N in the H burning ashes, 
   followed by double $\alpha$-capture on $^{14}$N. The residual $^{22}$Ne, 
   left behind at core He exhaustion, is fully consumed in the O-rich zone 
   during convective shell C burning. A small primary yield of Cu, 5 to 10 per 
   cent of its solar abundance depending on the location of the mass cut, 
   derives from explosive nucleosynthesis in the inner regions of 
   core-collapse supernovae (SNe\,II).

   In the early nineties, only 25 per cent of solar Cu was attributed to the 
   weak \emph{s-}process in massive stars (Raiteri et al. 1993). It was 
   advanced then (Matteucci et al. 1993; Mishenina et al. 2002) that most 
   solar Cu originated in type Ia supernovae (SNe\,Ia). Current SN\,Ia models, 
   however, predict a negligible production of Cu during thermonuclear 
   explosions (Iwamoto et al. 1999; Travaglio, Hillebrandt \& Reinecke 2005).

   It has been recently claimed that up-to-date, high-resolution spectroscopic 
   data for solar neighbourhood stars belonging to different Galactic 
   substructures (halo, thick-disc and thin-disc), bulge-like stars, globular 
   clusters (GCs) and dwarf spheroidal galaxies (dSphs) can all be well 
   understood if Cu is mostly a secondary product of massive stars, with a 
   small primary contribution from explosive nucleosynthesis (Bisterzo et al. 
   2005). While this conclusion simply rests on the comparison of extant 
   observations with present stellar nucleosynthesis (and Cu yields from stars 
   ultimately evolving to SNe\,II must still be viewed with great care -- see, 
   e.g., Heger et al. 2001; Hoffman, Woosley \& Weaver 2001; Rauscher et al. 
   2002), a detailed description of copper evolution in different galactic 
   structures is still missing from the literature.

   In this Letter, we try to fill in this gap by analysing the evolution of Cu 
   in the framework of detailed galactic chemical evolution (GCE) models in 
   which Cu originates mostly from (i) massive stars or (ii) SNe\,Ia. In 
   particular, we contrast the evolution of copper in the Milky Way with that 
   in the anomalous GC $\omega$\,Cen. We conclude that scenarios where copper 
   is produced primarily in SNe\,Ia are no longer supported by the 
   observations. This imposes strong constraints on the astrophysical 
   processes responsible for the synthesis of Cu in stars.

   \section{Models versus observations: Implications for stellar yields and 
            chemical evolution time scales}

%
   \begin{figure*}
   \psfig{figure=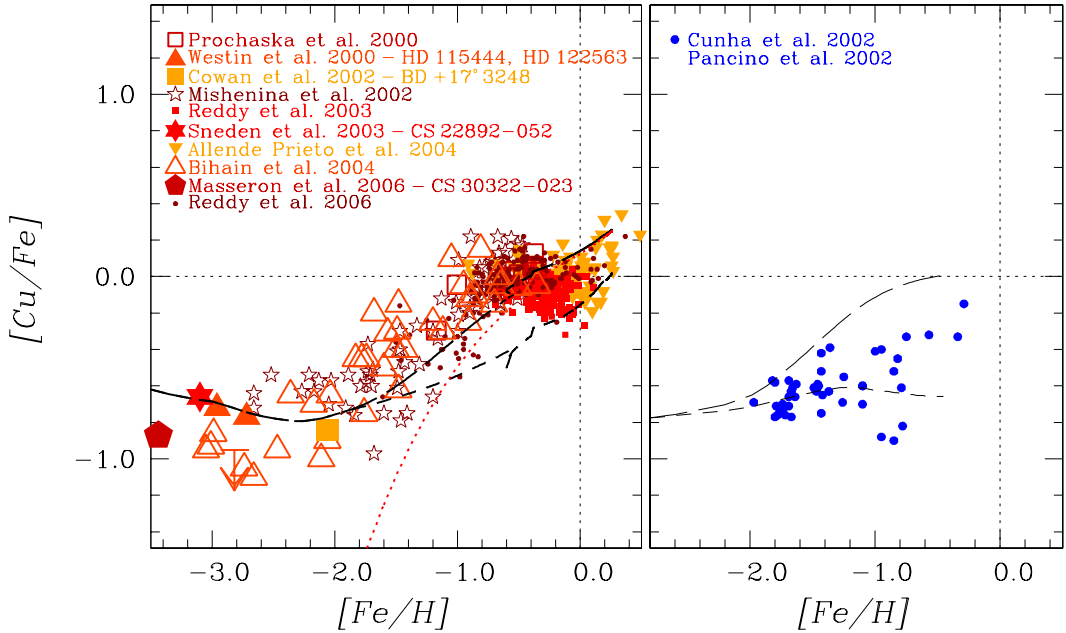,width=15cm} 
      \caption{ Left panel: [Cu/Fe] in Galactic disc and halo stars, as 
                measured by Prochaska et al. (2000; big empty squares), 
		Mishenina et al. (2002; stars), Reddy et al. (2003; small 
		filled squares), Allende Prieto et al. (2004; small filled 
		upside-down triangles), Bihain et al. (2004; big empty 
		triangles) and Reddy et al. (2006; dots) in the solar 
		neighbourhood. Also shown are [Cu/Fe] values for the very 
		metal-poor stars BD\,+17$^\circ$\,3248 (Cowan et al. 2002; big 
		filled square), HD\,115444, HD\,122563 (Westin et al. 2000; 
		big filled triangles) and the extremely metal-poor giant 
		CS\,22892$-$052 (Sneden et al. 2003; big star) and TP-AGB star 
		CS\,30322$-$023 (Masseron et al. 2006; pentagon). Right panel: 
		[Cu/Fe] in the globular cluster $\omega$\,Cen. Data from Cunha 
                et al. (2002) and Pancino et al. (2002; filled circles) have 
		been homogenized as discussed in Romano et al. (2007). 
		Superimposed on the data are our model predictions for 
		$\omega$\,Cen (right panel, thin lines) and for the solar 
		vicinity (left panel, thick lines), in cases where Cu 
		originates mostly from SN explosions (Models~1 and 2, Table~1, 
		long- and short-dashed lines, respectively; both type II and 
		type Ia SN explosions are considered). The effect of 
		completely removing the (primary) explosive SN\,II 
		contribution is also shown as a dotted line (only in a model 
		for the solar vicinity; left panel, Model~4, Table~1). See 
		text for details on different models' prescriptions.
              }
         \label{fig:cop1}
   \end{figure*}
%

   Nowadays, high-resolution spectroscopic data of [Cu/Fe] are available for 
   large samples of field Galactic stars (see Fig.~\ref{fig:cop1}, left 
   panel). GCs generally follow the trends defined by halo field giants 
   (Simmerer et al. 2003): a flat distribution, [Cu/Fe]~=~$-$0.75~$\pm$ 0.2, 
   for low-metallicity stars up to [Fe/H]~$\simeq$ $-$1.8 dex, followed by a 
   linear increase with a slope close to 1 in the metallicity range 
   $-$1.5~$<$~[Fe/H]~$<$ $-$1 (GC data are not plotted in 
   Fig.~\ref{fig:cop1}). At [Fe/H]~$>$ $-$0.8, [Cu/Fe] jumps above the solar 
   value. Then, a `bending' appears for disc stars with 
   $-$0.8~$<$~[Fe/H]~$<$~0 (Reddy et al. 2003), though there is a hint that 
   [Cu/Fe] might start increasing again at higher metallicities (Allende 
   Prieto et al. 2004).

\begin{figure*}
   \psfig{figure=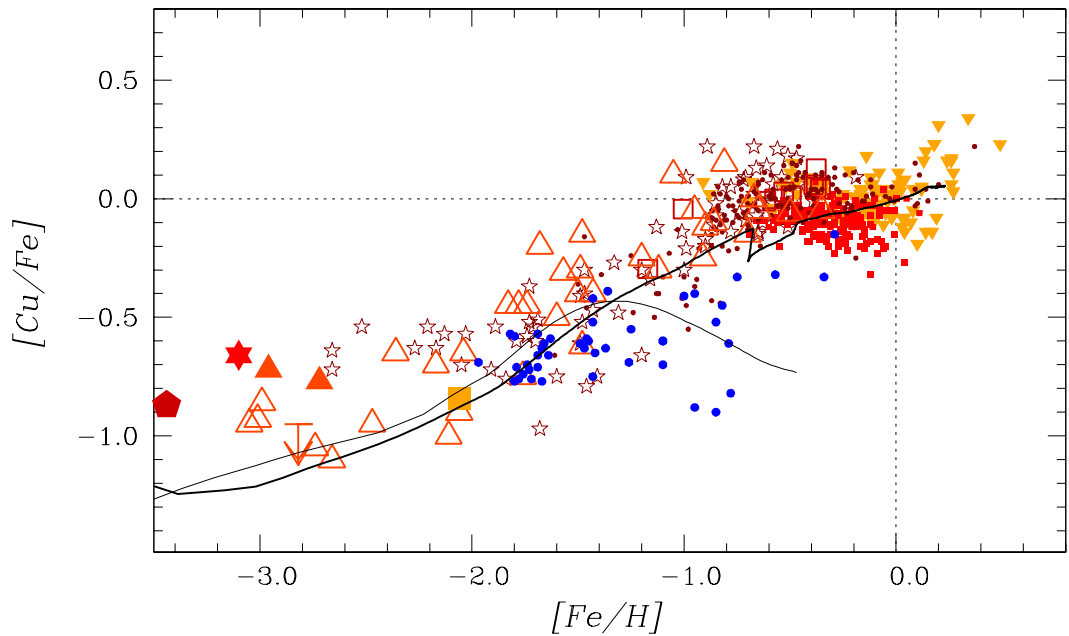,width=15cm} 
      \caption{ Same data as Fig.~\ref{fig:cop1}. The [Cu/Fe] versus [Fe/H] 
                relation in $\omega$\,Cen is now compared to that for solar 
		neighbourhood stars. Model predictions for $\omega$\,Cen (thin 
		solid line) and the solar vicinity (thick solid line) now 
		refer to the evolutive picture where most of solar Cu comes 
		from the weak \emph{s-}process in massive stars (stellar 
		nucleosynthesis prescriptions as in Model~3, Table~1). See 
		text for details.
              }
         \label{fig:cop2}
   \end{figure*}

   The Galactic GC $\omega$\,Cen stands as a notable exception. The [Cu/Fe] 
   ratios of its most metal-rich stars are definitely lower than the Galactic 
   trend (Cunha et al. 2002; Pancino et al. 2002; see also 
   Figs.~\ref{fig:cop1} and \ref{fig:cop2}), which can be understood as a 
   shift in the [Cu/Fe] relation to higher [Fe/H]. The only other systems 
   known to have unusually low [Cu/Fe] ratios are the Sagittarius dSph 
   (McWilliam \& Smecker-Hane 2005) and the Large Magellanic Cloud (LMC; 
   Pomp\'eia, Hill \& Spite 2005). In particular, the similarity between the 
   Cu values of $\omega$\,Cen and Sagittarius adds to the list of common 
   chemical peculiarities (spread in [Fe/H] values, strong enhancement of 
   \emph{s-}process elements) which suggest that the two systems followed 
   similar chemical enrichment histories, thus supporting the idea that 
   $\omega$\,Cen is the surviving nucleus of an accreted dwarf galaxy 
   (McWilliam \& Smecker-Hane 2005).

   Superposed on the data in Figs.~\ref{fig:cop1} and \ref{fig:cop2} are 
   different model predictions for $\omega$\,Cen (thin lines) and the Milky 
   Way (thick lines). These latter refer to the solar neighbourhood region 
   which, in the framework of the adopted model, is represented by a ring 
   2~kpc wide centred in the Sun. The model of chemical evolution for the 
   Galaxy is the \emph{two-infall model} of Chiappini, Matteucci \& Gratton 
   (1997) and Chiappini, Matteucci \& Romano (2001) -- where details about 
   model assumptions and basic equations can be found, except for the adopted 
   stellar lifetimes (which are now taken from Schaller et al. 1992) and 
   stellar yields (see the detailed discussion below). The model of chemical 
   evolution for $\omega$\,Cen is the one described in Romano et al. (2007), 
   assuming that $\omega$\,Cen is the remnant of an ancient nucleated dSph 
   evolved in isolation and then accreted by the Milky Way. We adopt this 
   simple model even if a recent study by Villanova et al. (2007) suggests 
   that the chemical evolution of $\omega$\,Cen might be more complex, with a 
   not unique age-metallicity relation. In fact, accounting for the observed 
   spread needs a full dynamical treatment, while the main conclusions on Cu 
   nucleosynthesis in stars are likely to remain the same.

%
%

%
   \begin{table*}
     \centering
       \caption{Prescriptions on stellar Cu production adopted by different 
                GCE models.}
       \begin{tabular}{@{}ccccl}
	 \hline
	 Model & M$^{\mathrm{ej}}_{\mathrm{Cu\, weak\, s}}$ & M$^{\mathrm{ej}}_{\mathrm{Cu\, SNeII}}$ & M$^{\mathrm{ej}}_{\mathrm{Cu\, SNeIa}}$ & Notes \\
	 \hline
	 1 & Matteucci et al. 1993 & Matteucci et al. 1993 (their model M) & Matteucci et al. 1993 (their model M) & Long-dashed lines in Fig.~\ref{fig:cop1}  \\
	 2 & Matteucci et al. 1993 & Matteucci et al. 1993 (their model M) & Iwamoto et al. 1999 (their model W7)  & Short-dashed lines in Fig.~\ref{fig:cop1} \\
	 3 & Nomoto et al. 2007    & Nomoto et al. 2007                    & Iwamoto et al. 1999 (their model W7)  & Continuous lines in Fig.~\ref{fig:cop2}   \\
	 4 & Matteucci et al. 1993 & 0                                     & Matteucci et al. 1993 (their model M) & Dotted line in Fig.~\ref{fig:cop1}        \\
	 \hline
       \end{tabular}
   \end{table*}
%

\begin{figure}
   \psfig{figure=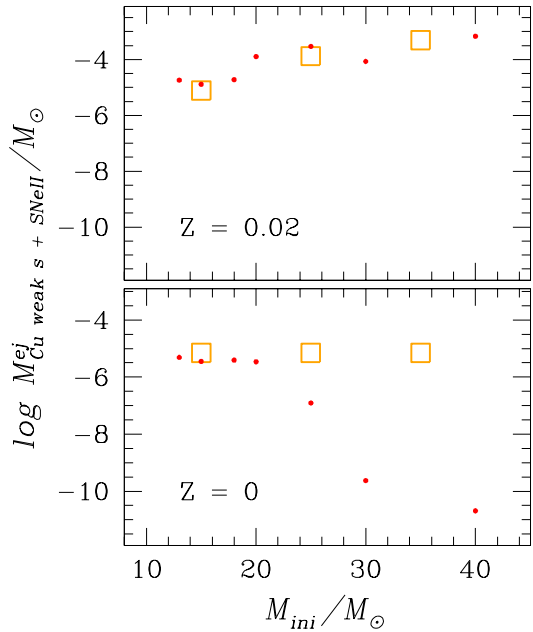,width=7.5cm} 
      \caption{ Stellar yields of Cu (in units of solar mass) for 
	        solar-metallicity and zero-metallicity massive stars from 
		Matteucci et al. (1993, squares) and Nomoto et al. (2007, 
		dots) as functions of the initial mass of the stars. At solar 
		metallicity (upper panel), the yields comprise a weak 
		\emph{s-}process component as well as an explosive one, while 
		for $Z$~= 0 (lower panel) only the explosive component is 
		present.
              }
         \label{fig:yields}
   \end{figure}

   As regards the adopted stellar nucleosynthesis:
   \begin{enumerate}
     \item the \emph{s-}process yields in massive stars vary as functions of 
           both the initial mass and metallicity of the stars; we follow the 
	   prescriptions of either Matteucci et al. (1993; Models~1, 2 and 4, 
	   Table~1; long-dashed, short-dashed and dotted lines in 
	   Fig.~\ref{fig:cop1}, respectively) or Nomoto et al. (2007; Model~3, 
	   Table~1; solid lines in Fig.~\ref{fig:cop2});
     \item explosive yields of Cu of individual core-collapse SNe are taken 
           from either Matteucci et al. (1993, their model~M, 
	   M$^{\mathrm{ej}}_{\mathrm{Cu\, SNeII}}$~= 7 $\times$ 10$^{-6}$ 
	   M$_\odot$, independent of mass and metallicity; Models~1 and 2, 
	   Table~1; long- and short-dashed lines in Fig.~\ref{fig:cop1}, 
	   respectively) or Nomoto et al. (2007; Model~3, Table~1; solid lines 
	   in Fig.~\ref{fig:cop2});
     \item the amount of Cu in the ejecta of SNe\,Ia is of 
           2~$\times$ 10$^{-4}$~M$_\odot$ (Matteucci et al. 1993, their 
	   model~M; Models~1 and 4, Table~1; long-dashed and dotted lines in 
	   Fig.~\ref{fig:cop1}) or two orders of magnitude lower (Iwamoto et 
	   al. 1999, their model~W7; Models~2 and 3, Table~1; short-dashed 
	   lines in Fig.~\ref{fig:cop1} and solid lines in 
	   Fig.~\ref{fig:cop2}), and it is independent of metallicity.
   \end{enumerate}
   Total yields of copper from massive stars by Matteucci et al. (1993, 
   squares) and Nomoto et al. (2007, dots), comprising both the weak component 
   and the explosive one, are shown and compared one another in 
   Fig.~\ref{fig:yields}, for solar metallicity stars (upper panel) and zero 
   metallicity objects (lower panel). At $Z$~= 0, the yields reflect the pure 
   explosive contribution. It is seen that, while the explosive yields adopted 
   by Matteucci et al. (1993) do not vary with the initial mass of the star, 
   those computed by Nomoto et al. (2007) vary steeply as a function of mass, 
   especially for $M_{\mathrm{ini}} >$ 20~M$_\odot$.

%
%

   Models where most of solar Cu comes from SNIa explosions following the 
   prescriptions of Matteucci et al. (1993) match very well the observational 
   data for solar neighbourhood stars (see Fig.~\ref{fig:cop1}, left panel, 
   thick long-dashed line), but leave the lower [Cu/Fe] ratios in 
   $\omega$\,Cen totally unexplained (see Fig.~\ref{fig:cop1}, right panel, 
   thin long-dashed line). When reducing the contribution from SNeIa according 
   to the results of Iwamoto et al. (1999), a better agreement with the 
   relation observed for $\omega$\,Cen members is found (except, perhaps, for 
   the two objects lying at the highest metallicities -- Fig.~\ref{fig:cop1}, 
   right panel, thin short-dashed line), but the agreement with the solar 
   neighbourhood data at [Fe/H]~$> -$1.0 is completely destroyed 
   (Fig.~\ref{fig:cop1}, left panel, thick short-dashed line). If, instead, 
   most of solar Cu is produced through the weak \emph{s-}process in massive 
   stars, both the solar neighbourhood and $\omega$\,Cen relations are 
   satisfactorily reproduced (Fig.~\ref{fig:cop2}, thick and thin solid 
   lines). It is worth stressing at this point that, in order to best fit the 
   solar abundance of Cu in the framework of our GCE model, we had to slightly 
   increase (by a factor of 1.5) the yields of Cu by Nomoto et al. (2007) for 
   solar metallicity stars. The applied correction factor, however, is well 
   inside current uncertainties on the weak \emph{s-}process modelling (see 
   Rauscher et al. 2002 for a discussion on this subject). It is also worth 
   emphasizing how dramatically important the production of Cu from 
   core-collapse SNe is in the very early evolution of the Galactic halo. The 
   SNII contribution sets the level of the `plateau' observed at [Fe/H]~$< 
   -$1.8. This can be clearly seen from Fig.~\ref{fig:cop1} (left panel), 
   where the thick dotted line (Model~4, Table~1) shows the effect of removing 
   any primary Cu production from explosive nucleosynthesis in massive stars 
   from the Milky Way model. In this case, the `plateau' at low [Fe/H] is no 
   longer expected, at variance with the observations, and only the steep rise 
   towards high [Cu/Fe] at high metallicities can be reproduced. Furthermore, 
   notice that when adopting the yields of Nomoto et al. (2007) rather than 
   those of Matteucci et al. (1993), a lower plateau level is obtained 
   (Fig.~\ref{fig:cop2}), because of the lower amount of Cu in the ejecta of 
   low-metallicity SNeII predicted by those authors (see 
   Fig.~\ref{fig:yields}, lower panel).

   Now we can draw the following conclusions:
   \begin{enumerate}
     \item After a short phase in which the primary contribution from 
           explosive nucleosynthesis in core-collapse SN dominates, the 
	   evolution of Cu in galaxies of different type is regulated mostly 
	   by the weak \emph{s-}process occurring in massive stars.
     \item Up-to-date, metallicity-dependent stellar yields of Cu by Nomoto et 
           al. (2007) including both a primary contribution from explosive 
	   nucleosynthesis and a secondary contribution from the 
	   \emph{s-}process in massive stars allow us to reproduce very well 
	   the observed behaviour of [Cu/Fe] versus [Fe/H] in the solar 
	   vicinity as well as in $\omega$\,Cen.
     \item When restricting our analysis to a single environment (the solar 
           vicinity) a degeneracy is found in the solution, in the sense that 
	   the decline in [Cu/Fe] with decreasing [Fe/H] can be ascribed to 
	   either the reduced extent of the weak \emph{s-}process in massive 
	   stars at low [Fe/H], or the delayed Cu production from SNeIa. Only 
	   studying objects which have chemically enriched along different 
	   evolutionary paths allows us to break the degeneracy.
   \end{enumerate}

%
   \begin{figure}
   \psfig{figure=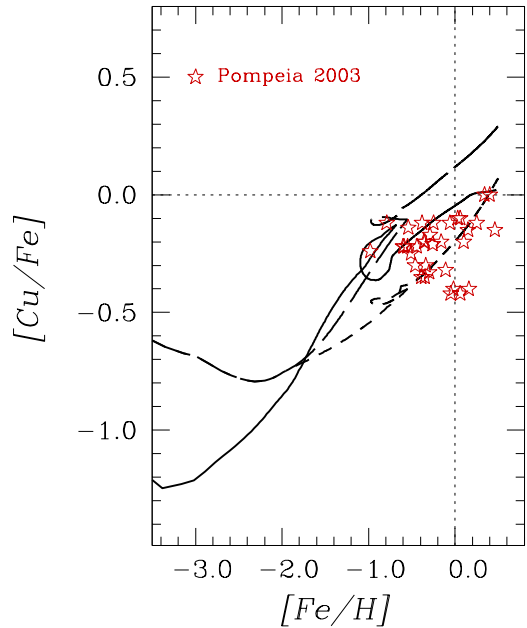,width=7.5cm} 
      \caption{ Predictions from Models~1 (long-dashed line), 2 (short-dashed 
	        line) and 3 (solid line) computed for the inner Galaxy 
		($R_{\mathrm{G}}$~= 4~kpc) compared to measurements of [Cu/Fe] 
		in a sample of bulge-like stars (data from Pomp\'eia 2003).
              }
         \label{fig:bls}
   \end{figure}
%

   Finally, in Fig.~\ref{fig:bls} we compare the abundances of Cu measured in 
   a sample of bulge-like stars by Pomp\'eia (2003) with our model predictions 
   for the inner Galaxy (bulge-like stars are thought to originate near the 
   Galactic bulge). A substantial contribution to the Galactic Cu production 
   from SNeIa (nucleosynthesis prescriptions as in Model~1, Table~1; 
   long-dashed line) is clearly ruled out, while models with a minimum Cu 
   production from SNeIa can well fit the data (Models~2 and 3, short-dashed 
   and solid lines, respectively). We conclude that the study of Cu abundances 
   \emph{in different environments,} from both an observational and a 
   theoretical point of view, is of fundamental importance in order to 
   understand the origin of Cu in the universe.

   \section*{Acknowledgments}
   Thanks are due to the referee, Raffaele Gratton, for his careful reading of 
   the manuscript and constructive comments.

\bsp

\label{lastpage}

\end{document}